\documentclass[10pt]{iopart}
\usepackage{graphicx}
\usepackage{iopams} 

\eqnobysec

\newcommand{\beq}{\begin{equation}}
\newcommand{\eeq}{\end{equation}}
\newcommand{\be}{\begin{equation*}}
\newcommand{\ee}{\end{equation*}}
\newcommand{\beqa}{\begin{eqnarray}}
\newcommand{\eeqa}{\end{eqnarray}}
\newcommand{\bea}{\begin{eqnarray*}}
\newcommand{\eea}{\end{eqnarray*}}

\def\stackunder#1#2{\mathrel{\mathop{#2}\limits_{#1}}}

\newcommand{\bigmean}[1]{\left\langle#1\right\rangle}

\newcommand{\bigac}[1]{\left(#1\right)}
\newcommand{\lap}[1]{\mathrel{\mathop{\cal L}\limits_{#1}^{}}}

\newcommand{\prob}{\mathop{\rm Prob}\nolimits}
\newcommand{\dd}{{\rm d }}

\renewcommand{\th}{{\theta}}

\renewcommand{\l}{\ell}

\newcommand{\stat}{\mathrm{stat}}

\newcommand{\taum}{\langle\tau\rangle}

\begin{document}

\title[Comment on `Fluctuation-dominated phase ordering at a mixed order transition' 
]{Comment on `Fluctuation-dominated phase ordering at a mixed order transition' 
}
\author{Claude Godr\`eche}
\address{
Institut de Physique Th\'eorique, Universit\'e Paris-Saclay, CEA and CNRS,
91191 Gif-sur-Yvette, France}\smallskip

\begin{abstract}
Renewal processes generated by a power-law distribution of intervals with tail index less than unity
are genuinely non-stationary.
This issue is illustrated by a critical review of the recent paper by Barma, Majumdar and Mukamel 2019
\textit{J.~Phys.~A} {\bf 52} 254001, devoted to the investigation of the properties of a specific one-dimensional equilibrium spin system with long-range interactions.
We explain why discarding the non-stationarity of the process underlying the model leads to an incorrect expression of the critical spin-spin correlation function,
even when the system, subjected to periodic boundary conditions, is translation invariant.
\end{abstract}

\section{Introduction}

Reference \cite{bmm} revisits the \textsc{tidsi} model, a truncated version of a microscopic one-dimensional spin model with long-range interactions, dubbed the inverse distance squared Ising (\textsc{idsi}) model \cite{idsi1,idsi2,idsi3}.
This \textsc{tidsi} model, which has been investigated in a series of papers in recent years \cite{bar1,bar2,bar3},
is 
made of a fluctuating number of domains, filling up the total size $L$ of the system.
A configuration is thus entirely specified by the number and sizes of these domains.
The Boltzmann weight associated to the realisation $\{\l_1,\l_2,\dots,\l_n,n\}$ of such a configuration 
reads \cite{bmm}
\beq\label{eq:bmm}
 p(\l_1,\l_2,\dots,\l_n,n|L)
=\frac{y^ng(\l_1)\dots g(\l_n)\delta\Big(\sum_{i=1}^n\l_i,L\Big)}{Z(L)},
\eeq
where the denominator is the partition function 
\beq\label{eq:Zbmm}
Z(L)=\sum_{n\ge1}\sum_{\l_1\dots\l_n} y^ng(\l_1)\dots g(\l_n)\delta\Big(\sum_{i=1}^n\l_i,L\Big)
\eeq
which ensures the normalisation, and $\delta(\cdot,\cdot)$ is the Kronecker delta.
In (\ref{eq:bmm}), $y$ denotes the fugacity and $g(\l)$ is given by
\beq\label{eq:gl}
g(\l)=\frac{1}{\l^{1+\th}}, 
\eeq
where $\l=1,2,\dots$, and the tail index $\th$ is positive.
In \cite{bmm,bar1,bar2} the phase diagram is analysed according to the values of the fugacity $y$ and the index $\th$, which are the two parameters of the model.
For $y=y_c$, such that
\beq\label{eq:yc}
y_c=\frac{1}{\sum_{\l\ge1} g(\l)}=\frac{1}{\zeta(1+\th)},
\eeq
where $\zeta(\cdot)$ is the Riemann zeta function, the system is critical, separating a paramagnetic (disordered) phase from a ferromagnetic (ordered) one.

\begin{figure}[!ht]
\begin{center}
\includegraphics[angle=0,width=.9\linewidth]{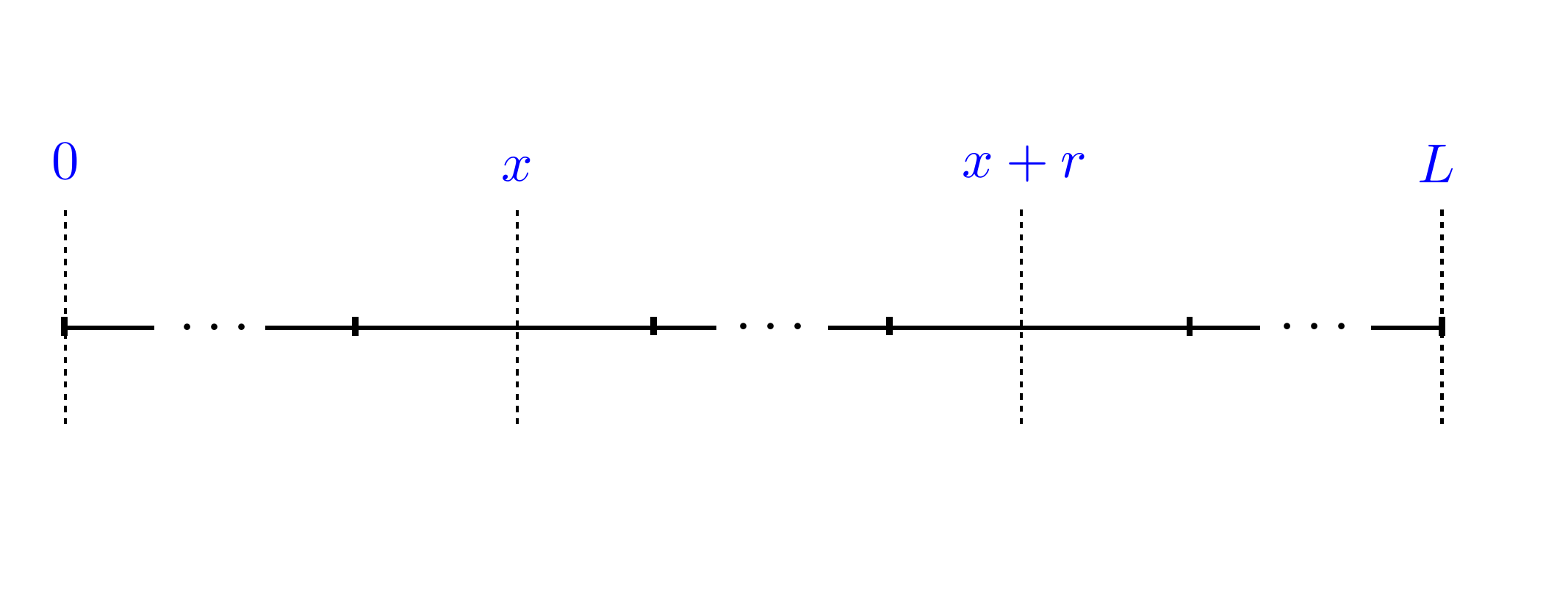}
\caption
{A tied-down renewal process is defined as the succession of a fluctuating number of time (or space) iid intervals between points (which can be events, domain walls, \dots.) conditioned to filling up the total size $L$ of the system.
The main quantity of interest in the present Comment is the probability $p_m(x,r|L)$ that $m$ points fall between the two arbitrary positions $x$ and $x+r$, and more particularly its expression for $m=0$, which is the probability that the interval $(x,x+r)$ does not contain any point.
}
\label{fig:corrSpin}
\end{center}
\end{figure}
The main goal of \cite{bmm} is to obtain the expression of the \textit{critical} correlation function between two spins located at $x$ and $x+r$, in the regime of short separations, i.e., where the separation $r$ between the two points is small compared to the system size $L$
 ($1\ll r\ll L$), when the tail index $\th<1$.
The expression derived in \cite{bmm} is
\beq\label{eq:bmmcorr}
G(r|L)\approx 1-A(\th)
\Big(\frac{r}{L}\Big)^{1-\th}, \qquad A(\th)=\frac{1}{\th}.
\eeq
As argued in \cite{bmm}, 
this expression is independent of the position $x$ of the first spin and only depends on the separation $r$,
because periodic boundary conditions are chosen, which makes the system translation invariant.

The principal aim of this Comment it to show that, in the present context where $\th<1$, the method used in \cite{bmm} in order to derive (\ref{eq:bmmcorr})---the so-called \textit{independent interval approximation}---does not lead to exact results, as claimed in this reference, and can only predict the scaling behaviour 
$G(r|L)-1\sim (r/L)^{1-\th}$ but not the amplitude $A(\th)$\footnote{Another method, similar in spirit, is also presented in \cite{bmm}, and leads again to the same result (\ref{eq:bmmcorr}).}. 
We shall also compare the study made in \cite{bmm} to the existing literature on the subject.
In this respect, we start by disproving an assertion made in \cite{bmm}, which contradicts a statement made in \cite{wendel2,wendel1}.

\section{The class of tied-down renewal processes}

As stated in \cite{wendel2,wendel1},
the \textsc{tidsi} model defined by (\ref{eq:bmm}), (\ref{eq:Zbmm}), (\ref{eq:gl}) belongs to the class of tied-down renewal processes
(independently of any considerations on the boundary conditions).

This can be very simply seen by introducing the parameter $w=y/y_c$ and making the change of notations
\beq\label{eq:tdc}
yg(\l)=wf(\l) \Longleftrightarrow f(\l)=y_c g(\l)=\frac{g(\l)}{\sum_{\l\ge1}g(\l)}.
\eeq
Now (\ref{eq:bmm}) and (\ref{eq:Zbmm}) respectively read
\beq\label{eq:jointTD}
 p(\l_1,\l_2,\dots,\l_n,n|L)
=\frac{w^nf(\l_1)\dots f(\l_n)\delta\Big(\sum_{i=1}^n\l_i,L\Big)}{Z(L)},
\eeq
and
\beq
\label{eq:Z}
Z(L)=\sum_{n\ge1}w^n\sum_{\l_1\dots\l_n} f(\l_1)\dots f(\l_n)\delta\Big(\sum_{i=1}^n\l_i,L\Big).
\eeq
Equation (\ref{eq:jointTD}) is nothing but the joint distribution for the class of tied-down renewal processes with a penalty or reward parameter $w$ (see \cite{cg2020} and references therein).
For this class of processes, a configuration is specified by a fluctuating number of intervals (e.g., domains)
$\tau_1,\tau_2,\dots$ 
which are independent and identically distributed (iid) random variables with common probability distribution 
\beq\label{eq:fldiscret}
f(\l)=\prob(\tau=\l).
\eeq
A realisation $\{\l_1,\l_2,\dots,\l_n,n\}$ of such a configuration, where the number of intervals takes the value $n$ and the random variables $\tau_i$ take the values $\l_i$, has weight (\ref{eq:jointTD}).
These processes are renewal processes because the intervals are
iid random variables, and they are tied down because these intervals are conditioned to sum up to a given value $L$,
 generalising the tied-down random walk, starting from the origin and conditioned to end at the origin at a given time \cite{wendel,wendel1}.
(For the tied-down random walk, intervals are temporal, while in the rest of this Comment all intervals are spatial.) 
The definitions (\ref{eq:jointTD}), (\ref{eq:Z}) and (\ref{eq:fldiscret}) easily generalise to the case where intervals are continuous random variables.

The parameter $w=y/y_c$ is larger than 1 in the disordered phase, equal to 1 at criticality, and less than 1 in the ordered phase.
The very same model---as defined by (\ref{eq:bmm}), (\ref{eq:Zbmm}) and (\ref{eq:gl})---was already introduced in \cite{burda}, where the phase diagram of the model was discussed.

The class of tied-down renewal processes defined by (\ref{eq:jointTD}), (\ref{eq:Z}) and (\ref{eq:fldiscret}) above, encompasses as a particular case the model defined in \cite{burda} and \cite{bmm,bar1,bar2,bar3}.
This class itself belongs to the broader class of \textit{linear models} described in \cite{fisher}, such as the Poland-Scherraga model, wetting models, etc.

To conclude, the following assertion, made in \cite{bmm}, does not hold true:
\textit{`A joint distribution similar to equation (10) was studied in the context of the tied-down renewal process [10], with the important difference that in the latter case the fugacity $y$ was taken to be exactly 1. 
As we will see later, in our \textsc{tidsi} model where $y$ can vary, there is a mixed-order phase transition at a critical value $y = y_c$ (which need not be 1)'}%
\footnote{In the first sentence quoted above, equation (10) refers to (\ref{eq:jointTD}), and reference [10] refers to \cite{wendel2}.}.
In these sentences, a confusion is made between the parameter $w$ and the fugacity $y$.
In the cited work [10] (see the footnote) the parameter taken to 1 is $w$ (hence $y=y_c$) and not the fugacity $y$.

From now on, $w$ is taken equal to 1, since all the present discussion concerns the \textit{critical} correlation function.
An important consequence of the above
is that all results found in \cite{wendel2,wendel1} hold for the \textsc{tidsi} model.
This applies, in particular, to (\ref{eq:taumcg}) and (\ref{eq:porod}) in section \ref{sec:recap}, as already stated in \cite{wendel2,wendel1}.

\section{Stationarity and non-stationarity of renewal processes}
\label{sec:recap}

Let us now give a short reminder of the 
relevant knowledge on two-space---or two-time---correlation functions in renewal theory, 
as a preparation for the critical review of \cite{bmm} given in section \ref{sec:bmm}.

\subsection{The correlation function}
\label{sec:corr}

Let $N(x,x+r)$ be the number of points (events, domain walls, \dots) comprised between two arbitrary positions $x$ and $x+r$ on the line, as depicted in figure \ref{fig:corrSpin}.
The correlation function 
of interest
is by definition 
\beq\label{eq:cxrL}
C(x,r|L)=\bigmean{(-1)^{N(x,x+r)}}=\sum_{m\ge0}(-1)^mp_m(x,r|L),
\eeq
where $p_m(x,r|L)$ is 
the probability distribution 
\beq\label{eq:pmxrL}
p_m(x,r|L)=\prob(N(x,x+r)=m).
\eeq
We are interested in the behaviour of this correlation function when the distribution $f(\l)$ of intervals is broad, with tail index $\th$, and tail parameter $c$,
\beq\label{eq:fltail}
f(\l)\stackunder{\l\to\infty}{\approx} \frac{c}{\l^{1+\th}},
\eeq
with emphasis on the case $\th<1$, and when $x, r,L$ are all simultaneously large.
In this regime, as shown in \cite{wendel2}, 
the correlation function 
$C(x,r|L)$ is dominated by $p_0(x,r|L)$, the probability that the interval $(x,x+r)$ does not contain any point.
From now on we shall focus on this quantity.
We assume that the first interval begins at site $0$.
For the sake of simplicity we use a continuum formalism for the random intervals $\tau_i$.
In the sequel, we will need the expansion of the Laplace transform of $f(\l)$,
\beq\label{eq:hatfs}
\hat{f}(s)\stackunder{{s\rightarrow 0}}{\approx}\left\{ 
\begin{array}{ll}
1+a\,s^{\th}+\cdots, & (\th <1) \\ 
1-\left\langle \tau\right\rangle s+\cdots+a\,s^{\th}+\cdots, & (\th>1),
\end{array}
\right. \qquad \label{ro_broad}
\eeq
where 
\beq\label{eq:a}
a=c\,\Gamma (-\th)
\eeq
is negative if $\th<1$, positive if $1<\th<2$, and so on.

For \textit{free renewal processes}, there is no conditioning at $L$, which is equivalent to taking the limit $L\to\infty$ in the expressions found for the tied-down case, as will be checked in
\S \ref{sec:corrRenew}.
The definition of the correlation function is thus still given by (\ref{eq:cxrL}), however now with no reference to $L$.

\subsection{Correlation function for tied-down renewal processes}
\label{sec:corrTD}
The correlation function $C(x,r|L)\approx p_0(x,r|L)$ has several regimes according to the respective magnitudes of $x$, $r$, and $L$ \cite{wendel2}.
Here we restrict the presentation to the regime of short separations, where 
$1\ll r\ll x\sim L$, which is the only regime considered in \cite{bmm}.

According to the nature of the distribution $f(\l)$, the following dichotomy holds \cite{wendel1}.

\medskip\noindent 
\textbf{1. $\taum$ finite} 

If the distribution of intervals $f(\l)$ has a finite first moment
\beq\label{eq:taumdef}
\taum=\int_0^\infty\dd \l\, \l f(\l),
\eeq
e.g., when $f(\l)$ is narrow, with finite moments, or broad with a power-law tail of index $\th>1$,
then, when
$1\ll r\ll x\sim L$, the system enters a stationary regime, where $p_0$ no longer depends on $x$ and $L$ as can be derived from the analytical expression in Laplace space of this observable \cite{wendel2}.
Its asymptotic expression is the same as in the free case and is given in (\ref{eq:freep0Short}).

Let us denote by $\taum^\star$ the first moment of the marginal size distribution of a generic interval, $f(\l |L)$, obtained by tracing the full distribution 
$p(\l_1,\l_2\cdots,\l_n,n|L)$ (\ref{eq:jointTD}) on all $\l_i$ but one.
Alternatively, $\taum^\star$ is equal to the product of $L$ by the mean inverse number of intervals in $(0,L)$.
Then, if $\th>1$, at large $L$, $\taum^\star$ is asymptotically equal to $\taum$ \cite{wendel1}.

\medskip\noindent 
\textbf{2. $\taum$ infinite} 

The situation is different when $f(\l)$ is a broad distribution with tail index $\th<1$.
Then $\taum$ is no longer finite, which is the source of non-stationarity,
 and \cite{wendel1}
\beq\label{eq:taumcg}
\taum^\star\approx\frac{\pi c}{\sin\pi\th}\,L^{1-\th},
\eeq
where $c$ is the tail parameter of the distribution $f(\l)$ (see (\ref{eq:fltail})).
In the \textsc{tidsi} model for example, $c$ is equal to $y_c$ (see (\ref{eq:gl}) and (\ref{eq:tdc})).

In the regime of \textit{short separations} between $x$ and $x+r$ ($1\ll r\ll x\sim L$), the correlation function, $C(x,r|L)\approx p_0(x,r|L)$, has the following expression \cite{wendel2}
\beqa\label{eq:porod}
C(x,r|L)
&\approx &
1-\frac{\sin\pi\theta}{\pi(1-\theta)}\left[\frac{x}{L}\left(1-\frac{x}{L}\right)\right]^{\theta-1}
\left(\frac{r}{L}\right)^{1-\theta}
\nonumber\\
&\approx&1-\frac{\sin\pi\theta}{\pi(1-\theta)}\left(\frac{rL}{x(L-x)}\right)^{1-\theta}.
\eeqa
This expression is non-stationary, since it depends on the ratios $r/L$ and $x/L$, and universal since it only depends on $\th$ and no longer on microscopic details of the distribution $f(\l)$ such as the tail parameter $c$.

\subsection{Correlation function for free renewal processes}
\label{sec:corrRenew}
Now, the correlation function is a function of $x$ and $r$ only, and we still have $C(x,r)\approx p_0(x,r)$ for large values of the arguments, with the dependence in $L$ dropped out in the notation.
The dichotomy seen in the previous subsection still holds.

\medskip\noindent 
\textbf{1. $\taum$ finite} 

In this situation, in the limit $x\to\infty$,
the process reaches a stationary regime, where $p_0(x,r)\to p_{0,\stat}(r)$ (see, e.g., \cite{cox,gl2001}).
In Laplace space with respect to $r$, with conjugate variable $s$,
\beq\label{eq:lapp0r}
\lap{r}p_{0,\stat}(r)=\hat p_{0,\stat}(s)=\frac{1}{s}-\frac{1-\hat f(s)}{\taum s^2}.
\eeq
For $r$ large, we use the second line of the expansion (\ref{eq:hatfs}) for $s\to0$, which yields, by inversion, the stationary result \cite{gl2001},
\beq\label{eq:freep0Short}
p_{0}(x,r)\stackunder{1\ll r\ll x}{\approx} p_{0,\stat}(r)\approx
\frac{c}{\th(\th-1)\taum}r^{-(\th-1)}.
\eeq

\medskip\noindent 
\textbf{2. $\taum$ infinite}

When $\taum$ is infinite, $p_0(x,r)$ keeps a dependence in $x$, even at large values of this variable, which is the signature of the non-stationarity of
the process \cite{gl2001,bouchaud}.
For $x$ and $r$ simultaneously large and comparable, $p_0(x,r)$ has the scaling form
\beq\label{eq:gscaling}
p_0(x,r)\approx g\Big(\frac{x}{x+r}\Big),
\eeq
where the universal scaling function $g(\cdot)$ reads \cite{gl2001}
\beq\label{eq:gscaling+}
g(\xi)=\int_0^{\xi}\dd u\, \frac{u^{\th-1}(1-x)^{-\th}}{\Gamma(\th)\Gamma(1-\th)}.
\eeq
In the regime of short separations, (\ref{eq:gscaling}) and (\ref{eq:gscaling+}) yield
\beq\label{eq:shortR}
p_0(x,r)\stackunder{1\ll r\ll x}{\approx} 1-\frac{\sin \pi\th}{\pi(1-\th)}\Big(\frac{r}{x}\Big)^{1-\th}.
\eeq
As announced earlier, (\ref{eq:shortR}) is precisely the $L\to\infty$ limit of (\ref{eq:porod}).

\section{Derivation of (\ref{eq:bmmcorr}) in the independent interval approximation}
\label{sec:bmm}

We now come to the derivation of (\ref{eq:bmmcorr}) given in \cite{bmm}, when $\th<1$, by the so-called \textit{independent interval approximation}.
A similar approach has already been used in \cite{das1,das2}.

\subsection{IIA method}
\label{sec:IIA}

The method, as presented in \cite{bmm}, proceeds as follows.
\begin{enumerate}
\item The system is taken infinite ($L$ is sent to infinity), \textit{stationary} and the distribution of the sizes of domains $f(\l)$ is assumed to have a finite first moment $\taum$.
Therefore the stationary probability that an interval of size $r$ does not contain any point is 
given by (\ref{eq:lapp0r}) in Laplace space, which expresses $\hat p_{0,\stat}(s)$ in terms of the Laplace transform $\hat f(s)$ and $\taum$.
\item In this formalism, $f(\l)$ is thought as being the marginal $f(\l|L)$, but nevertheless
approximated by $f(\l)$, if $\l\ll L$,
and
$\taum$ is thought as being $\taum^{\star}$, and given, according to \cite{bmm}, by
\beq\label{eq:taumstar}
\taum^\star\approx \frac{c}{1-\th} L^{1-\th},
\eeq
which allows to reintroduce $L$ in the formalism.
\item
The expansion of the Laplace transform of $f(\l)$ with respect to $\l$, valid if $\th<1$,
\beq\label{eq:expansion}
\hat f(s)\approx 1+a s^\th,
\eeq
where $a=c\,\Gamma(-\th)$, see (\ref{eq:a}),
is carried into the expression (\ref{eq:lapp0r}) of $\hat p_{0,\stat}(s)$.
This gives
\be
\hat p_{0,\stat}(s)\approx \frac{1}{s}+\frac{c\,\Gamma(-\th)}{\taum}s^{\th-2}.
\ee
Now $\taum$ is replaced by $\taum^{\star}$ given by (\ref{eq:taumstar}),
yielding finally, after inversion, 
\beq\label{eq:p0bmm}
p_{0,\stat}(r)\approx 1-\frac{1}{\th}\bigac{\frac{r}{L}}^{1-\th},
\eeq
which is (\ref{eq:bmmcorr}), since the correlation function is dominated by $p_{0,\stat}(r)$.
\end{enumerate}

\subsection{Discussion}

According to \cite{bmm}, choosing periodic boundary conditions for the system, entailing translation invariance,
justifies the use of a stationary formalism where the dependence in $x$ is discarded at the very start, instead of stemming naturally from an analytical computation.
In reality, as analysed below, the treatment given in \cite{bmm} cannot lead to an exact prediction for the correlation function, when $\th<1$.

The range of validity of the IIA method is summarised in point (i) above.
If consistently completed---in particular by using the second line of the expansion (\ref{eq:hatfs}) of $\hat f(s)$---it would lead to (\ref{eq:freep0Short}) for the correlation function, which has not the expected form 
$G(r|L)-1\sim (r/L)^{1-\th}$.
Additional assumptions are therefore introduced in points (ii) and (iii).
These assumptions unfortunately 
 do not form a coherent whole.
On the one hand, using (\ref{eq:lapp0r}) requires $\taum$ to be finite.
On the other hand, using the expansion (\ref{eq:expansion}) only makes sense if $\taum$ is infinite.
If $\taum$ is finite one has to use the expansion in the second line of (\ref{eq:hatfs}).
Taking $\taum^\star$ in place of $\taum$ does not circumvent this contradiction because the finiteness of $\taum$ is an intrinsic property of the distribution $f(\l)$, independent of the size of the system---i.e., holding even for an infinite system.
In contrast, $\taum^\star$ is a property of the finite system and depends on $L$.

The IIA method actually does not know about the boundary conditions because it consists of the formal analysis of an infinite system, which can neither take account of a finite size, nor of boundary conditions.
This is reflected in the comparison between \cite{bmm} with the arXiv version \cite{arxiv} of the same work.
Except for the change in boundary conditions from \textit{open} in \cite{arxiv}, to \textit{periodic} in \cite{bmm}, all computations made in \cite{arxiv} remained unchanged, leading to the same result (\ref{eq:bmmcorr}).

To sum up, the IIA is used in \cite{bmm} outside its range of validity.
It only predicts the power $1-\th$ of $r/L$ in (\ref{eq:bmmcorr}) or (\ref{eq:p0bmm}).
It has no predictive value for the amplitude $A(\th)$ in (\ref{eq:bmmcorr}) or (\ref{eq:p0bmm}).
All the more as the expression (\ref{eq:taumstar}) for $\taum^\star$ is inaccurate and should be replaced by (\ref{eq:taumcg}).
By doing so, the amplitude in (\ref{eq:bmmcorr}) or (\ref{eq:p0bmm}) would be changed to 
$\sin\pi\th/[\pi\th(1-\th)]$,
which does not suffice to give the correct expression for the correlation function either.

\section{The role of boundary conditions}

To the three possible geometries,
\begin{enumerate}
\item the infinite line, 
\item the finite system of length $L$ as in figure \ref{fig:corrSpin},
\item the circle of length $L$,
\end{enumerate}
correspond three different expressions of the correlation function, assuming $\th<1$.

For the infinite line, the result is given by (\ref{eq:shortR}) in the short-separation regime (which is the only regime discussed in \cite{bmm} and in this Comment).
For the finite system of length $L$, the result is given by (\ref{eq:porod}).
These expressions are asymptotic estimates of exact results \cite{gl2001,wendel2}.

Remains to predict an expression for the last case (iii).
The question boils down to finding the probability that an interval of size $r$ located anywhere on a circle of length $L$ contains zero point.
If the interval is $(x,x+r)$, since $x$ can be anywhere between 0 and $L$, i.e., is uniformly distributed on the circle, one has to integrate uniformly upon this variable the expression (\ref{eq:porod}) of the probability that there is zero point in the interval $(x,x+r)$.
One thus obtains
\beq\label{eq:porodPBC}
G(r|L)
\approx 
1-\frac{\sin\pi\theta}{\pi(1-\theta)}\frac{\Gamma(\th)^2}{\Gamma(2\th)}
\left(\frac{r}{L}\right)^{1-\theta}.
\eeq
See \cite{arxivcg,cgtocome} for more details.

\section{Conclusion}

Tied-down renewal processes generated by a power-law distribution of intervals with tail index $\th<1$,
of which the critical \textsc{tidsi} model (a spin model with long-range order) is a particular example, are genuinely non-stationary.
For such processes, the independent interval approximation put forward in \cite{bmm} for the computation of the spin-spin correlation function is not the proper approach, 
because it is applied outside its range of validity and is based on the formal analysis of an infinite system which can neither handle a finite system, nor boundary conditions.
When periodic boundary conditions are imposed on the system, entailing translation invariance, the correlation function of the system is obtained by tracing the non-stationary correlation function $C(x,r|L)$ uniformly on $x$, resulting in (\ref{eq:porodPBC}).

\end{document}